\pdfoutput=1
\documentclass[10pt,conference]{IEEEtran}

\usepackage{cite}
\usepackage{amsmath,amssymb,amsfonts}
\usepackage{graphicx}
\usepackage{array}
\usepackage{textcomp}
\usepackage{xcolor}
\makeatletter
\@ifpackageloaded{hyperref}{%
  \hypersetup{hidelinks,bookmarks=false,bookmarksopen=false}%
}{%
  \usepackage[hidelinks,bookmarks=false,bookmarksopen=false]{hyperref}%
}
\makeatother
\usepackage{booktabs}
\usepackage{placeins}
\usepackage{float}

\graphicspath{{./}}

\newcommand{\maybeincludegraphics}[2]{%
    \IfFileExists{#2}{%
        \includegraphics[width=#1]{#2}%
    }{%
        \fbox{%
            \parbox[c][0.22\textheight][c]{0.95#1}{%
                \centering Missing figure\\ \texttt{\detokenize{#2}}%
            }%
        }%
    }%
}

\begin{document}

\title{Diffusion-Based Stress Testing of Overload Monitoring for Resilient Emergency Cellular Networks Using Internet CDR Proxies}

\author{
\IEEEauthorblockN{Bilal Hussain\IEEEauthorrefmark{1}, 
Xiao Tang\IEEEauthorrefmark{2}, 
Tan Li\IEEEauthorrefmark{3},
Muhammad Azhar\IEEEauthorrefmark{4},
Danista Khan\IEEEauthorrefmark{1}, and Fawad Ahmad\IEEEauthorrefmark{1}}
\IEEEauthorblockA{\IEEEauthorrefmark{1}College of Professional and Continuing Education, The Hong Kong Polytechnic University, Hong Kong, China}
\IEEEauthorblockA{\IEEEauthorrefmark{2}School of Information and Communication Engineering, Xi'an Jiaotong University, Xi'an, China}
\IEEEauthorblockA{\IEEEauthorrefmark{3}Department of Computer Science, The Hang Seng University of Hong Kong, Hong Kong, China}
\IEEEauthorblockA{\IEEEauthorrefmark{4}Department of Applied Data Science, Hong Kong Shue Yan University, Hong Kong, China}
}

\maketitle

\begin{abstract}
Disasters can overload cellular control-plane signaling within minutes, yet fine-grained Radio Resource Control (RRC) or Next Generation (NG) Application Protocol (NGAP) telemetry is privacy-sensitive and costly to collect for analytics. Many emergency monitoring pipelines therefore rely on coarse Call Detail Record (CDR) aggregates. We treat Internet activity in CDR grids as a practical proxy for hidden signaling stress under that constraint. We train a lightweight convolutional neural network (CNN) on stylized overload injections, stress-test it with diffusion-synthesized surges that preserve normal traffic structure, and adapt the detector by retraining on hard synthetic samples. Under stress-test conditions, the default alert threshold fails even though receiver operating characteristic (ROC) curves stay strong: the detector still assigns overloaded cells a larger overload probability than normal cells, but those probabilities fall below the default cutoff $0.5$ and are labeled normal, so the F1-maximizing threshold---selected post hoc on the same stress-test grids (oracle $\tau^*$)---shifts by $0.32 \pm 0.03$ (operating-point drift). Across three random seeds, hard-sample adaptation raises thresholded performance (F1) from $0\%$ (no alerts at the default cutoff $0.5$ on any seed) to $85.67 \pm 14.37$\% and ranking from ROC-AUC $0.886 \pm 0.040$ to $0.99996 \pm 0.00007$. Diffusion-synthesized surges expose threshold fragility that matched-condition training---training and testing on the same stylized injections---hides, and hard-sample adaptation restores usable alerts at the default cutoff. Together, these steps define a reusable pre-deployment stress test for emergency monitors. Internet-only CDR input further supports lightweight AI-native workflows that combine monitoring, recalibration, and adaptation.
\end{abstract}


\section{Introduction}
\label{sec:introduction}

Earthquakes, floods, evacuations, and mass-casualty incidents can overload control-plane signaling long before user-plane capacity fails~\cite{khan2025storm,nguyen2025rrc}, degrading cellular service within minutes when coordination matters most. The November 2025 Wang Fuk Court fire in Tai Po, Hong Kong---which killed 168 people, displaced nearly 5{,}000 residents, and led a subsequent public inquiry to report that territory-wide mobile emergency alerts can take up to one hour to reach users---illustrates why alert and overload monitoring must remain dependable during urban disasters~\cite{scmp2026taipofire}. Emergency operators therefore need rapid, coarse-grained stress indicators so that slicing, traffic steering, aerial backhaul, and priority signaling can activate before collapse~\cite{saad2025agi,zhang2026mahead}.

Fine-grained control-plane telemetry (e.g., RRC or NGAP message traces) is privacy-sensitive and costly to collect for security measurement~\cite{He2018}, so many analytics pipelines instead use Call Detail Record (CDR) aggregates~\cite{parwez2017bigdata}. Aggregated CDR views support cell-level spatial monitoring~\cite{naboulsi2016large,sultan2018call}, but no single CDR channel is a direct measurement of signaling overload. Following our earlier CDR-grid formulation, in which elevated Internet activity approximates a signaling storm~\cite{hussain2021ddos}, we therefore treat Internet CDR activity as an \emph{indirect} proxy for that hidden control-plane stress when only such coarse aggregates are available. The Milano CDR collection~\cite{barlacchi2015dataset} supplies the real cell-level traffic grids on which we learn normal behavior and construct labeled synthetic surges for this study.

\begin{figure*}[t]
\centering
\maybeincludegraphics{0.78\textwidth}{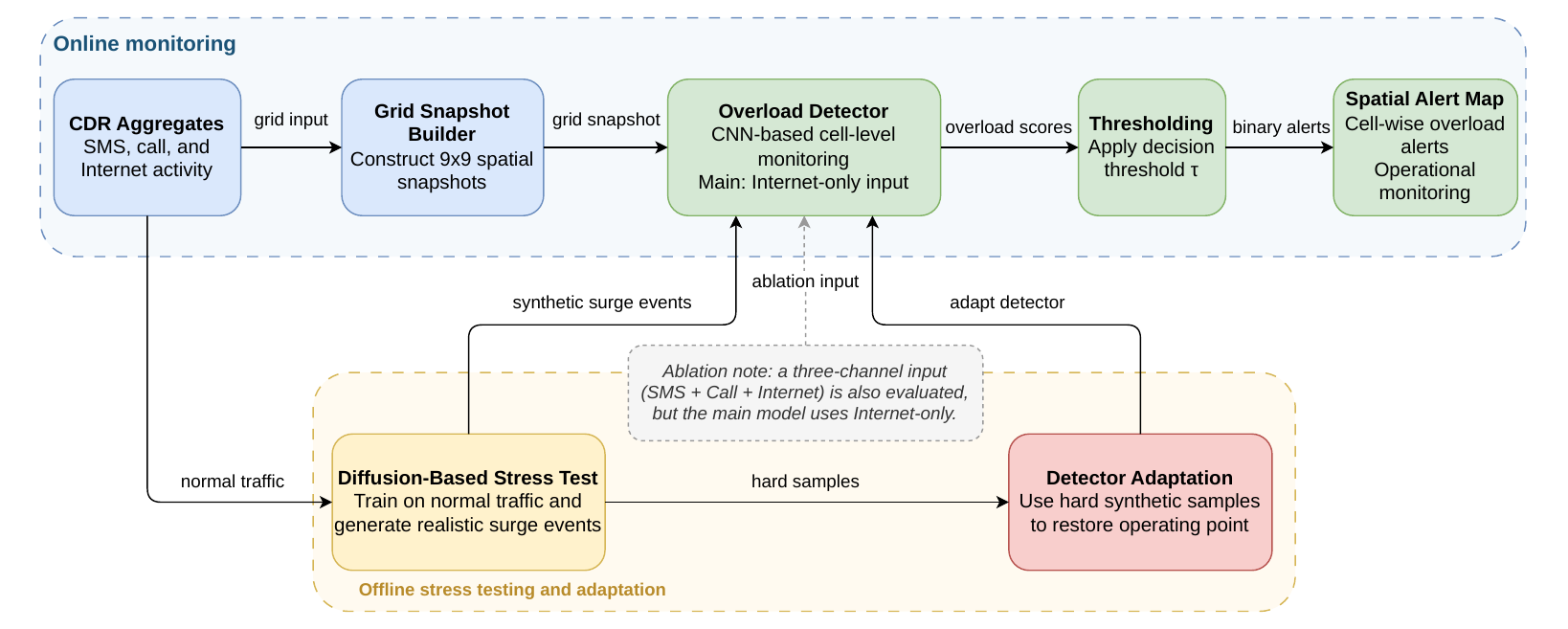}
\vspace{-7pt}
\caption{Emergency overload monitoring framework. \emph{Online} (top): $9 \times 9$ snapshots scored by a compact CNN and thresholded at fixed $\tau=0.5$. \emph{Offline} (bottom): diffusion stress testing and hard-sample adaptation; $\tau^*$ is selected offline on stress-test grids and is not the online cutoff. Main detector: Internet-only; ablation: three-channel input.}
\label{fig:system_model}
\end{figure*}

A second challenge is evaluation realism. Detectors trained and tested on the same stylized overload injections can look reliable yet fail under realistic surge shift. Diffusion models can generate harder, structure-preserving traffic samples~\cite{ho2020ddpm,kotelnikov2023tabddpm,zhang2025semantic}. We train on stylized Internet-channel surges, stress-test with diffusion-synthesized events, and adapt using difficult synthetic samples. Whereas prior CDR-grid studies often emphasize grid representation or channel-aligned detection~\cite{hussain2021ddos,sultan2018call,aziz2024tmc}, this paper emphasizes emergency overload monitoring under distribution shift, explicit threshold-drift analysis, and post-shift adaptation.

This work complements our companion IEEE Transactions on Industrial Informatics (TII) paper~\cite{hussain2026diffddos}, which evaluates multi-scenario malicious-traffic detection (Short Message Service (SMS), silent-call, Internet, blended) with guided diffusion retraining and fast gradient sign method (FGSM), projected gradient descent (PGD), and conditional tabular generative adversarial network (CTGAN) baselines in a CPS-security setting. Relative to that study and the CDR-grid multiplier construction in~\cite{hussain2021ddos}, this paper adds Internet-only emergency-overload framing, operating-point drift $\Delta\tau$, multi-seed thresholded F1, and calibration-only versus hard-sample adaptation on a lightweight single-channel detector---not a re-run of the TII attack-suite benchmarks.

We aim to (i) formulate Internet CDR proxies for signaling-overload monitoring; (ii) stress-test fixed-threshold detectors with diffusion-synthesized surges; and (iii) quantify recovery through $\Delta\tau$, adaptation, and calibration. Our contributions are:
\begin{itemize}\setlength{\itemsep}{0pt}\setlength{\parsep}{0pt}\setlength{\topsep}{2pt}
    \item An emergency-oriented signaling-overload monitoring formulation using Internet CDR proxies.
    \item A diffusion-based stress-test protocol that exposes threshold fragility under realistic distribution shift.
    \item A resilience interpretation based on operating-point drift and post-adaptation recovery, with multi-seed evaluation.
    \item A comparison of \texttt{internet\_only} (compact CNN) versus \texttt{three\_channel} (ResNet50) input--architecture pairs on this proxy task.
    \item A calibration-versus-adaptation comparison and diffusion realism validation that address matched-condition over-optimism in emergency monitoring evaluation.
\end{itemize}

Relative to next-generation intelligent wireless emergency communications, this study supports \emph{network planning and scheduling} through operating-point drift analysis and threshold recalibration before overload actions, and contributes a reusable \emph{testing platform} through the diffusion-based stress-test protocol for evaluating monitors under disaster-like surge shift.

Section~\ref{sec:related} reviews related work; Sections~\ref{sec:method}--\ref{sec:results} present methods and results; Section~\ref{sec:discussion} discusses limitations; Section~\ref{sec:conclusion} concludes.

\section{Related Work}
\label{sec:related}

\textbf{Signaling overload and disaster-driven surges.} Signaling storms and control-plane overload threaten modern cellular networks when attach, paging, and session-management activity overwhelms core and radio access network (RAN) resources~\cite{khan2025storm,nguyen2025rrc,park2022signaling}. Emergency surges resemble these patterns even without intentional traffic manipulation~\cite{cazzolato2023tgrapp,lakhina2004diagnosing}. Monitors must remain useful under naturally occurring shift, not only under stylized laboratory conditions.

\textbf{CDR-based cellular analytics.} CDR aggregates are widely used for anomaly detection, traffic forecasting, and cell-level network analytics~\cite{naboulsi2016large,sultan2018call,aziz2024tmc}, and public corpora such as Milano enable reproducible studies~\cite{barlacchi2015dataset}. Prior CDR-grid studies usually monitor the channel that directly reflects the event (SMS flooding in SMS counts, data overload in Internet counts)~\cite{hussain2021ddos,sultan2018call,aziz2024tmc}. In this paper the target event is control-plane signaling stress. Because fine-grained Radio Resource Control (RRC) or NG Application Protocol (NGAP) telemetry is often costly to expose for analytics, Internet activity in CDR aggregates remains a practical observable for cell-level overload monitoring. Relative to those channel-aligned detectors, we evaluate the Internet-proxy monitor under distribution shift, treat operating-point drift ($\Delta\tau$) as a resilience metric, compare adaptation against threshold-only recalibration, and check diffusion-surge realism.

\textbf{Diffusion models for synthetic traffic.} Denoising diffusion probabilistic models (DDPMs)~\cite{ho2020ddpm} and the tabular extension TabDDPM~\cite{kotelnikov2023tabddpm} generate realistic low-dimensional traffic samples while preserving aggregate structure; diffusion also supports harder out-of-distribution sample generation~\cite{nie2022diffusion}. Matched train--test evaluation on stylized surges can overstate deployment reliability~\cite{arp2022dos}, so this paper uses diffusion as an emergency-network evaluation platform and threshold-calibration probe rather than only as a data-augmentation tool. Our companion TII work~\cite{hussain2026diffddos} applies the same idea to multi-scenario malicious-traffic robustness in the context of CPS security.

\section{Methodology}
\label{sec:method}

We study cell-level overload monitoring using CDR aggregates under three evaluation stages: \textbf{Stage~1 / baseline} (matched stylized surges), \textbf{Stage~2 / stress-test} (diffusion-synthesized surges), and \textbf{Stage~3 / adaptation} (retraining on hard synthetic samples), as summarized in Fig.~\ref{fig:system_model}.

\subsection{Problem Setting and System Model}
\label{subsec:problem_setting}

Each observation is a grid snapshot $\mathbf{X} \in \mathbb{R}^{9 \times 9 \times C}$, where $C$ is the number of CDR channels. The main detector uses \texttt{internet\_only} input ($C{=}1$) with a compact CNN (three convolution--pooling blocks, $8/16/32$ filters, $256$-unit dense layer, $81$ sigmoids). A \texttt{three\_channel} comparison ($C{=}3$) uses ResNet50: $9 \times 9$ grids are zero-padded by $5$ pixels to $19 \times 19$ before the $7 \times 7$ stem, then residual stages and an $81$-way sigmoid head. The two settings also change the backbone, so they are input--architecture pairs, not a controlled channel-only ablation. Binary labels $\mathbf{y} \in \{0,1\}^{81}$ mark each cell as normal ($0$) or overload-affected ($1$); because the Milano CDR corpus provides no overload ground truth, these labels are assigned by the synthetic surge construction in Sec.~\ref{subsec:dataset_scope}. The detector outputs cell-wise overload probabilities $\hat{\mathbf{y}} \in [0,1]^{81}$.

Fig.~\ref{fig:system_model} separates an online monitoring path (CDR aggregates to $9 \times 9$ grid snapshots, compact CNN scores, thresholding at fixed $\tau=0.5$, and a spatial alert map) from an offline path (diffusion-based stress testing, detector adaptation, and oracle $\tau^*$ on stress-test grids). The lightweight single-channel design targets disaster-response settings where full three-channel processing may be unavailable.

\subsection{Proxy Validity Rationale}
\label{subsec:proxy_validity}

Fine-grained control-plane telemetry is privacy-sensitive and costly to collect for security measurement~\cite{He2018}, so practical analytics often operate on CDR aggregates~\cite{parwez2017bigdata}. Internet activity recorded in CDRs captures data-session bursts that co-occur with control-plane overload during disasters and signaling storms~\cite{khan2025storm,nguyen2025rrc,park2022signaling}, although they do not measure signaling directly. We evaluate whether an Internet-proxy monitor still assigns higher overload probabilities to overloaded cells than to normal cells under shift and whether recalibration or adaptation restores usable alerts---without claiming one-to-one equivalence to RRC load without paired signaling traces.

\subsection{Dataset Scope, Grid Construction, and Baseline Surges}
\label{subsec:dataset_scope}
\label{subsec:grid_construction}

We evaluate on the public Milano CDR corpus (2013--2014)~\cite{barlacchi2015dataset}, which reflects 3G/4G-era metropolitan traffic rather than a 5G standalone core capture. We do not claim that decade-old traffic statistics match modern 5G loads; rather, coarse per-cell service aggregates remain a generation-agnostic observability layer for operator-side monitoring~\cite{aziz2024tmc}, and disaster-driven signaling overload remains relevant in contemporary 5G/Open Radio Access Network (O-RAN) deployments~\cite{khan2025storm,nguyen2025rrc}. Aziz and Bestak~\cite{aziz2024tmc} argue that although individual CDR fields evolve across radio-access generations, the core record format and service-count semantics remain comparable, so CDR-based analytics can extend forward without redesigning the monitoring representation.

Following~\cite{hussain2021ddos,hussain2026diffddos}, Milano is retained as a reproducibility benchmark so that diffusion stress testing and operating-point drift analysis are isolated from dataset changes; the empirical question is whether fixed-threshold monitors fail under realistic surge shift on the CDR aggregate manifold and whether adaptation restores usable alerts. To our knowledge, no public labeled per-cell cellular overload CDR corpus offers a drop-in replacement, so we learn normal traffic from real CDR, synthesize labeled surge grids, and stress-test monitors under distribution shift, consistent with established CDR-grid practice~\cite{sultan2018call,hussain2021ddos}. Grid construction follows the CDR-grid preprocessing pipeline in~\cite{hussain2021ddos} and our companion IEEE TII diffusion study~\cite{hussain2026diffddos}; snapshot counts, temporal window, and train/test splits match Table~\ref{tab:setup}.

\begin{table}[!t]
\caption{Experimental setup for the emergency overload study.}
\label{tab:setup}
\centering
\footnotesize
\begin{tabular}{>{\raggedright\arraybackslash}p{0.28\columnwidth} >{\raggedright\arraybackslash}p{0.60\columnwidth}}
\toprule
\textbf{Item} & \textbf{Value} \\
\midrule
Dataset & Telecom Italia Big Data Challenge~\cite{barlacchi2015dataset} \\
Scenario & Internet activity as signaling-overload proxy \\
Grid, cells & $9 \times 9$, 81 \\
Window, slots & 11:00--14:00, 18 per day \\
Days or files & 62 \\
Normal train, test & 614, 167 \\
Stress train, test & 167, 168 \\
Overload-affected cells & 40 \\
\bottomrule
\end{tabular}
\end{table}

To obtain labeled baseline overload grids, we select a fixed overload-affected cell subset $\mathcal{C}$ ($|\mathcal{C}|=40$) and scale only the Internet activity of those cells by a constant multiplier $\alpha$:
\begin{equation}
X_{k,\text{Internet}} \leftarrow \alpha X_{k,\text{Internet}}, \qquad k \in \mathcal{C},
\end{equation}
where the left arrow denotes in-place replacement of the original Internet count in cell $k$, and $\alpha{=}43.3$ follows the Internet-signaling emulation derived in the appendix of~\cite{hussain2021ddos}. We set $y_k=1$ for $k \in \mathcal{C}$ and $y_k=0$ otherwise, so overload is defined operationally by this synthesized Internet surge, not by measured RRC. Normal grids are unperturbed Milano snapshots (Table~\ref{tab:setup}). This stylized surge is strongly separated from normal Internet aggregates (${\approx}59\sigma$), making matched-condition training diagnostically easy. Training uses weighted multilabel binary cross-entropy on the backbone assigned to each input mode in Sec.~\ref{subsec:problem_setting}.

\subsection{Diffusion Stress-Test Generator}
\label{subsec:stress_generator}

To move beyond stylized surges, the Diffusion-Based Stress Test block in Fig.~\ref{fig:system_model} fits a tabular diffusion model on normal 3D CDR aggregates $\mathbf{x}_0 = [x_{\text{SMS}}, x_{\text{Call}}, x_{\text{Internet}}] \in \mathbb{R}^{3}$. Following DDPM~\cite{ho2020ddpm}, the forward noising process is
\begin{equation}
q(\mathbf{x}_t \mid \mathbf{x}_0) =
\mathcal{N}\!\left(
\sqrt{\bar{\alpha}_t}\mathbf{x}_0,\,
(1-\bar{\alpha}_t)\mathbf{I}
\right),
\end{equation}
where $t$ is the diffusion step and $\bar{\alpha}_t$ is the cumulative noise-schedule coefficient. A denoiser $\epsilon_\psi$ is trained with the standard noise-prediction objective
\begin{equation}
\mathcal{L}_{\mathrm{DDPM}} =
\mathbb{E}_{\mathbf{x}_0,t,\epsilon}\left[
\left\|
\epsilon - \epsilon_\psi(\mathbf{x}_t,t)
\right\|_2^2
\right],
\end{equation}
with $\mathbf{x}_t = \sqrt{\bar{\alpha}_t}\mathbf{x}_0 + \sqrt{1-\bar{\alpha}_t}\,\epsilon$ and $\epsilon \sim \mathcal{N}(\mathbf{0},\mathbf{I})$. At sampling time, the learned reverse process generates realistic aggregate rows on the normal CDR manifold. Tabular design choices follow TabDDPM~\cite{kotelnikov2023tabddpm}; CDR-specific instantiation details are deferred to~\cite{hussain2026diffddos}.

Synthetic surge rows are drawn from the learned normal sampler and perturbed only on Internet activity by a milder random multiplier $\gamma$:
\begin{equation}
x_{\text{Internet}} \leftarrow \gamma x_{\text{Internet}}, \quad \gamma \text{ drawn uniformly from } [2.0, 5.0],
\end{equation}
then mapped back into $\mathcal{C}$ on a background grid. Relative to the fixed baseline multiplier $\alpha{=}43.3$ (${\approx}59\sigma$), these milder Internet shifts (${\approx}3.6\sigma$) are used as near-manifold emergency-like stress tests (Sec.~\ref{subsec:realism_validation}), not as paired disaster traces. This procedure is used only during stress-test evaluation.

\subsection{Adaptation with Hard Synthetic Samples}
\label{subsec:detector_adaptation}

Adaptation corresponds to the Detector Adaptation block in Fig.~\ref{fig:system_model} and applies a lightweight variant of the guided retraining procedure in~\cite{hussain2026diffddos}: the diffusion sampler generates candidate surge events, and detector feedback steers sampling toward hard yet overload-like examples. Those hard samples are added to the training set, and the detector is fine-tuned for a small number of epochs to stabilize monitoring under shifted conditions~\cite{xiao2023secure}.

\subsection{Threshold and Resilience Metrics}
\label{subsec:threshold_analysis}

Because realistic surges can shift score distributions~\cite{guo2017calibration}, overload probabilities on affected cells may remain higher than on normal cells, yielding a high area under the receiver operating characteristic curve (ROC-AUC), yet still fall below a fixed cutoff such as $\tau=0.5$. Every cell is then labeled normal and F1---the harmonic mean of precision and recall on micro-averaged cell decisions---collapses at the Thresholding stage in Fig.~\ref{fig:system_model}. We therefore report both ROC-AUC and threshold-dependent F1 after sweeping $\tau \in \{0.05,0.10,\ldots,0.95\}$. As a diagnostic for fixed-threshold emergency deployment, we additionally define the \emph{operating-point drift magnitude}
\begin{equation}
\Delta\tau = \left|\tau_{\mathrm{default}} - \tau_{\mathrm{F1max}}\right|,
\end{equation}
where $\tau_{\mathrm{default}}=0.5$ is the online cutoff and $\tau_{\mathrm{F1max}}$ (also denoted $\tau^*$) is the F1-maximizing threshold on a given evaluation set. Reported $\tau^*$ and $\Delta\tau$ are \emph{oracle} (post-hoc) values on those grids, not independently validated operating points. When reporting multi-seed results we use the mean $\pm$ std of per-seed $\tau^*$ values. Large $\Delta\tau$ indicates that the default alert rule has become fragile under surge shift.

\subsection{Evaluation Protocol}
\label{subsec:evaluation_protocol}

The protocol contains three stages: \textbf{(i) Stage~1 (baseline evaluation)} on matched stylized surges along the online detector path in Fig.~\ref{fig:system_model}; \textbf{(ii) Stage~2 (stress-test evaluation)} on diffusion-synthesized surge events (Diffusion-Based Stress Test); and \textbf{(iii) Stage~3 (adaptation evaluation)} after retraining on hard synthetic samples (Detector Adaptation). Multi-seed runs vary initialization, shuffling, and overload-affected cell selection.

For the \textbf{calibration-only baseline}, we keep the baseline-trained detector fixed and select $\tau_{\mathrm{F1max}}$ by sweeping $\tau$ on the same stress-test grids used for reporting (oracle; no held-out calibration split and no weight updates). All reported metrics use this multi-seed configuration: seeds $1$--$3$ (a workshop-scale budget, not a safety-certification sample); detector training for $24$ epochs; TabDDPM with $T{=}300$ and $30$ training epochs; and guided hard-sample retraining for $2$ rounds with $64$ samples/round and $12$ fine-tuning epochs/round.

\section{Results}
\label{sec:results}

Unless noted otherwise, results average three seeds. Snapshot counts and splits follow Table~\ref{tab:setup}.

\subsection{Baseline and Stress-Test Behavior}
\label{subsec:baseline_stress}

\begin{figure}[!t]
\centering
\maybeincludegraphics{0.72\columnwidth}{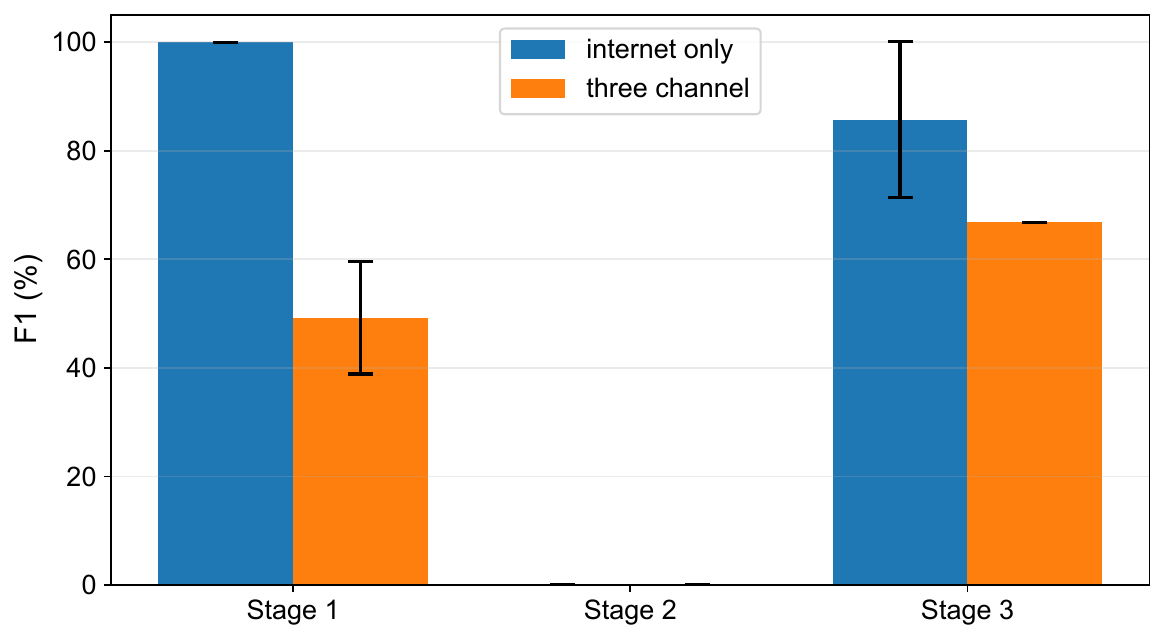}
\vspace{-5pt}
\caption{Multi-seed F1 by evaluation stage (Stage~1: baseline; Stage~2: stress-test; Stage~3: adaptation). Error bars show one standard deviation across three runs.}
\label{fig:multiseed_f1}
\end{figure}

Under baseline conditions, \texttt{internet\_only} reaches $99.94 \pm 0.05$\% F1 (Table~\ref{tab:main_results}, Fig.~\ref{fig:multiseed_f1}), so the lightweight single-channel detector looks near-perfect when train and test surges match. Operationally, a lightweight monitor would pass validation if checked only under matched laboratory-style surges. The three-channel pair stays much weaker ($49.21 \pm 10.36$\%): the stylized surge scales only Internet activity, so SMS and call channels remain near-normal distractors and dilute the proxy signature even under matched conditions.

\begin{table}[!t]
\caption{Multi-seed summary across Stage~1--3 evaluation (baseline, stress-test, and adaptation). Abbrev.: Eval., evaluation.}
\label{tab:main_results}
\centering
\resizebox{\columnwidth}{!}{%
\begin{tabular}{llccc}
\toprule
\textbf{Mode} & \textbf{Eval.} & \textbf{Runs} & \textbf{F1 Mean $\pm$ Std (\%)} & \textbf{ROC-AUC Mean $\pm$ Std} \\
\midrule
internet\_only & Baseline & 3 & 99.94 $\pm$ 0.05  & $-^{\,\mathrm{a}}$ \\
internet\_only & Stress-test & 3 & 0.00  $\pm$ 0.00  & 0.886 $\pm$ 0.040 \\
internet\_only & Adaptation & 3 & 85.67 $\pm$ 14.37 & 0.99996 $\pm$ 0.00007 \\
three\_channel & Baseline & 3 & 49.21 $\pm$ 10.36 & $-^{\,\mathrm{a}}$ \\
three\_channel & Stress-test & 3 & 0.00  $\pm$ 0.00  & 0.264 $\pm$ 0.033 \\
three\_channel & Adaptation & 3 & 66.80 $\pm$ 0.00$^{\,\mathrm{b}}$ & 0.964 $\pm$ 0.059 \\
\bottomrule
\end{tabular}%
}\\[2pt]
\parbox{\columnwidth}{\footnotesize
    \textsuperscript{a}~Baseline ROC-AUC\,$\approx$\,1.0; omitted as uninformative.\\
    \textsuperscript{b}~Mixed-set F1; stress-only F1 is 100\% (Section~\ref{subsec:adaptation}).
}
\end{table}

The stress-test stage tells a different story. At $\tau=0.5$, \texttt{internet\_only} F1 collapses to $0.00 \pm 0.00$\%, yet ROC-AUC remains $0.886 \pm 0.040$ (Table~\ref{tab:main_results}). The mean oracle F1-maximizing threshold $\tau^*$ shifts to $0.18 \pm 0.03$ (Fig.~\ref{fig:threshold_sweep}, left panel, Stage~2), giving $\Delta\tau = 0.32 \pm 0.03$. An apparently failed monitor can still assign higher overload probabilities to overloaded cells than to normal cells while needing immediate threshold recalibration; fixed $\tau=0.5$ rules risk false negatives during disaster surges even when the model stays informative.

The \texttt{three\_channel} pair degrades further under stress-test conditions (ROC-AUC $0.264 \pm 0.033$), with $\Delta\tau=0.45$. That larger drift signals broader score miscalibration and higher false-alarm risk if deployed without careful calibration during emergency response.

\subsection{Diffusion Sample Realism}
\label{subsec:realism_validation}

Table~\ref{tab:realism_validation} compares Internet-proxy surges from our lightweight TabDDPM checkpoint ($T{=}300$, $30$ training epochs). We extract one 3D aggregate vector per stress-test grid (mean over $|\mathcal{C}|{=}40$ overload-affected cells) and compare it against $5{,}000$ normal cell-level references using target-channel mean $|z|$ and radial basis function (RBF) maximum mean discrepancy (MMD)$^2$, following~\cite{hussain2026diffddos}. TabDDPM stress samples deviate by $3.61\sigma$ (joint MMD$^2{=}0.877$), whereas the $\alpha{=}43.3$ multiplier deviates by $58.76\sigma$, explaining matched-condition over-optimism~\cite{arp2022dos}. As a coarse spatial check, Internet-channel patterns on stress-test grids remain weakly correlated with the mean normal spatial template ($0.219 \pm 0.146$ for TabDDPM vs.\ $0.248 \pm 0.117$ for multipliers); fine-grained spatial realism remains future work.

\begin{table}[!t]
\caption{Realism of Internet-proxy surge construction (168 stress-test grids; TabDDPM, $T{=}300$).}
\label{tab:realism_validation}
\centering
\footnotesize
\begin{tabular*}{\columnwidth}{@{\extracolsep{\fill}}lrrr@{}}
\toprule
\textbf{Surge source} & \textbf{$|z|$} & \textbf{MMD$^2_{\mathrm{I}}$} & \textbf{MMD$^2_{\mathrm{3D}}$} \\
\midrule
TabDDPM + $\gamma{\in}[2,5]$ & 3.61 & 0.874 & 0.877 \\
Multiplier ($\alpha{=}43.3$) & 58.76 & 0.603 & 0.602 \\
TabDDPM raw ($5k$ draws) & --- & --- & 0.509 \\
\bottomrule
\end{tabular*}\\[1pt]
\parbox{\columnwidth}{\scriptsize Recomputed on the pipeline checkpoint used in this study; $|z|$ on Internet; MMD$^2_{\mathrm{I}}$ on Internet; MMD$^2_{\mathrm{3D}}$ on joint 3D aggregates.}
\end{table}

\subsection{Input Representation Comparison}
\label{subsec:input_comparison}

\begin{figure}[!t]
\centering
\maybeincludegraphics{0.90\columnwidth}{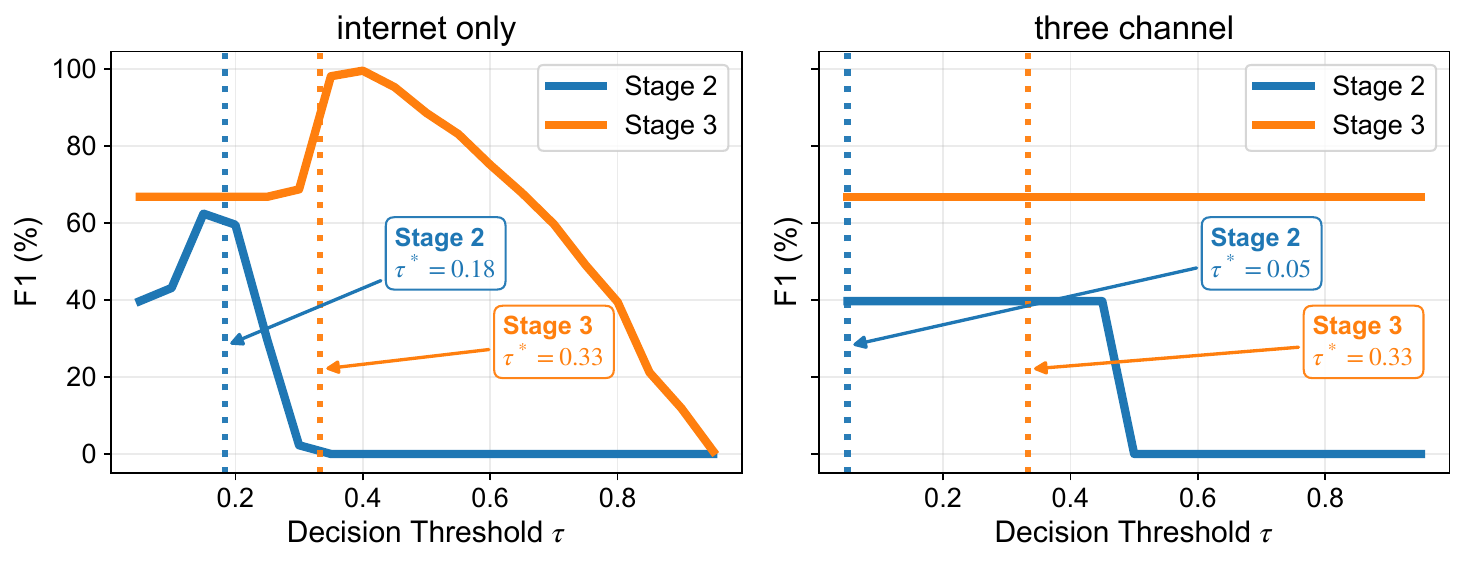}
\vspace{-7pt}
\caption{Threshold sweep under the Internet overload-proxy scenario (left: \texttt{internet\_only}; right: \texttt{three\_channel}). Curves show representative single-run F1--$\tau$ sweeps; dotted lines (annotated) mark the multi-seed mean of per-seed oracle F1-maximizing thresholds $\tau^*$ for Stage~2 and Stage~3.}
\label{fig:threshold_sweep}
\end{figure}

Table~\ref{tab:ablation_results} summarizes the input--architecture comparison. The compact-CNN \texttt{internet\_only} pair dominates every stress-relevant metric: higher stress-test AUC ($0.886$ vs.\ $0.264$), higher post-adaptation F1 ($85.67$ vs.\ $66.80$), and smaller stress-test drift ($\Delta\tau=0.317$ vs.\ $0.450$); stress-test F1 is $0.00$ for both modes at $\tau{=}0.5$. Because backbone and channels change together, we do not isolate a pure channel effect. Fig.~\ref{fig:threshold_sweep} shows F1 versus $\tau$ (predict overload if score $\ge\tau$; left: \texttt{internet\_only}; right: \texttt{three\_channel}): Internet-only recovers a wider usable threshold band after adaptation, whereas three-channel stays brittle at low $\tau$ under stress-test. Dotted $\tau^*$ lines are multi-seed means of per-seed oracle F1-maxima, not necessarily the peak of the plotted single-run curve.

\begin{table}[!t]
\caption{Input--architecture comparison and operating-point drift. Abbrev.: St.-test, stress-test; Adapt., adaptation; AUC, area under the curve.}
\label{tab:ablation_results}
\centering
\resizebox{\columnwidth}{!}{%
\begin{tabular}{lcccc}
\toprule
\textbf{Mode} & \textbf{St.-test AUC} & \textbf{Adapt. F1} & \textbf{Adapt. AUC} & \textbf{$\Delta\tau$ (St., Ad.)} \\
\midrule
internet\_only & 0.886 & 85.67 & 0.99996 & 0.317, 0.167 \\
three\_channel & 0.264 & 66.80 & 0.96421 & 0.450, 0.167 \\
\bottomrule
\end{tabular}%
}\\[2pt]
\parbox{\columnwidth}{\footnotesize
    Stress-test F1 is $0.00$ for both modes at $\tau{=}0.5$ (Table~\ref{tab:main_results}). $\Delta\tau$ uses oracle $\tau^*$ on evaluation grids. Modes differ in both input and backbone (compact CNN vs.\ ResNet50).
}
\end{table}

When only coarse CDR summaries can run locally, monitoring the Internet channel alone reduces compute load while yielding clearer overload signatures for this proxy task.

\subsection{Recovery After Adaptation}
\label{subsec:adaptation}

After adaptation, \texttt{internet\_only} recovers to $85.67 \pm 14.37$\% F1 and $0.99996 \pm 0.00007$ ROC-AUC. The F1 standard deviation is large for a safety-critical claim: ranking is near-perfect, but thresholded F1 varies across three seeds, so $\tau=0.5$ F1 is not a certified operating point. Oracle $\tau^*$ moves to $0.33 \pm 0.06$ ($\Delta\tau=0.17$; Table~\ref{tab:ablation_results}; Fig.~\ref{fig:threshold_sweep}, Stage~3).

\begin{figure}[!t]
\centering
\maybeincludegraphics{0.68\columnwidth}{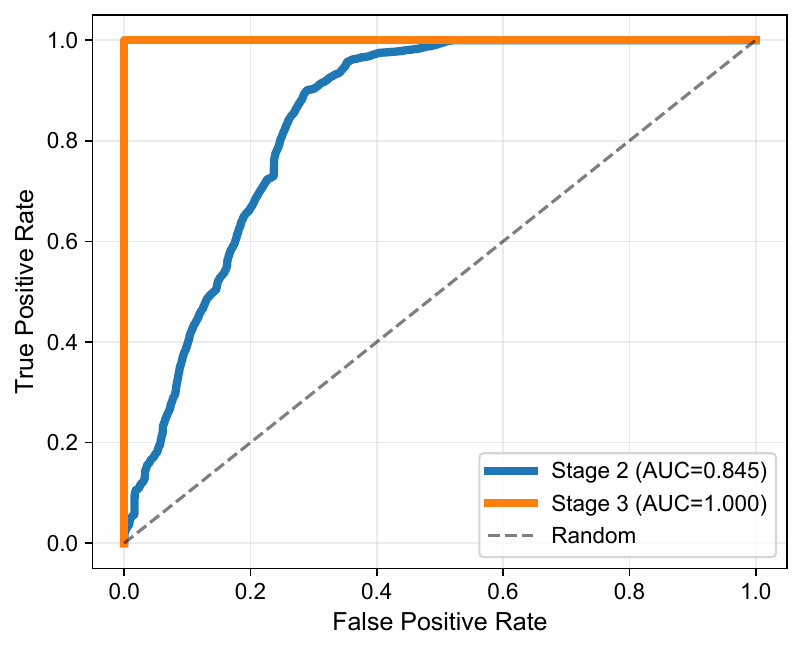}
\vspace{-7pt}
\caption{ROC curves before and after adaptation for \texttt{internet\_only} (representative single-run curves). Multi-seed mean AUC values appear in Table~\ref{tab:main_results}.}
\label{fig:roc_proxy}
\end{figure}

Fig.~\ref{fig:roc_proxy} shows that adaptation restores strong separability and narrows the gap between ranking and decision quality. For emergency workflows, this supports a two-step response: detect drift under stress-test conditions, then retrain or recalibrate before the next incident window.

The three-channel model attains 100\% F1 on stress-only samples after adaptation but only $66.80 \pm 0.00$\% on the mixed normal-plus-stress set. During emergency response, that specificity failure can trigger unnecessary rerouting or priority-service actions over large urban areas. Under this protocol, the compact Internet-only CNN is the stronger of the two input--architecture pairs.

\subsection{Calibration-Only Baseline vs.\ Adaptation}
\label{subsec:calibration_baseline}

Table~\ref{tab:calibration_baseline} compares recovery strategies under stress-test shift. Threshold-only recalibration ($\tau_{\mathrm{F1max}}$, no retraining) raises F1 from $0.00 \pm 0.00$\% to $67.07 \pm 5.48$\%, while hard-sample adaptation reaches $85.67 \pm 14.37$\% at $\tau{=}0.5$ and $98.95 \pm 1.19$\% with post-adaptation $\tau_{\mathrm{F1max}}$. Adaptation improves both ranking (AUC $\rightarrow 0.99996$) and thresholded performance beyond calibration alone.

\begin{table}[!t]
\caption{Calibration-only baseline vs.\ hard-sample adaptation on stress-test grids (\texttt{internet\_only}, three seeds).}
\label{tab:calibration_baseline}
\centering
\footnotesize
\setlength{\tabcolsep}{3pt}
\begin{tabular}{>{\raggedright\arraybackslash}p{0.30\columnwidth} c r r}
\toprule
\textbf{Strategy} & \textbf{$\tau$} & \textbf{F1 (\%)} & \textbf{AUC} \\
\midrule
Fixed threshold & 0.5 & 0.00{\scriptsize$\pm$0.00} & 0.886{\scriptsize$\pm$0.04} \\
Threshold only & $\tau^*$ & 67.07{\scriptsize$\pm$5.48} & 0.886{\scriptsize$\pm$0.04} \\
Adaptation & 0.5 & 85.67{\scriptsize$\pm$14.37} & 0.99996{\scriptsize$\pm$0.00007} \\
Adapt. + threshold & $\tau^*$ & \textbf{98.95{\scriptsize$\pm$1.19}} & \textbf{0.99996{\scriptsize$\pm$0.00007}} \\
\bottomrule
\end{tabular}\\[1pt]
\parbox{\columnwidth}{\scriptsize $\tau^*$: oracle F1-maximizing threshold on the same stress-test grids used for reporting; no retraining for rows 1--2.}
\end{table}

\subsection{Spatial Consistency}
\label{subsec:spatial_consistency}

\begin{figure}[!t]
\centering
\maybeincludegraphics{0.68\columnwidth}{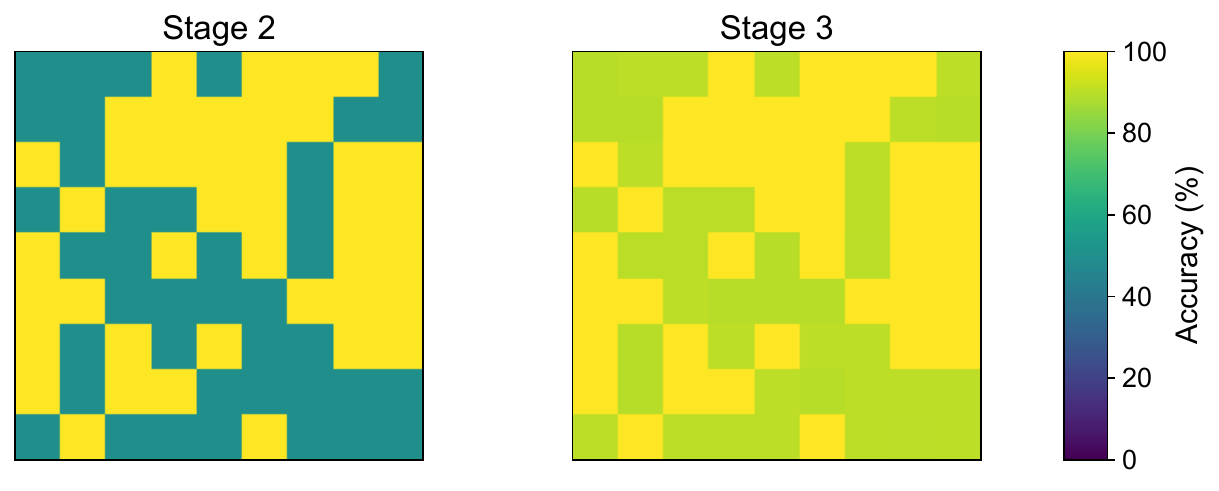}
\vspace{-7pt}
\caption{Per-cell accuracy heatmaps under diffusion-synthesized surges before (left) and after (right) adaptation.}
\label{fig:heatmap_proxy}
\end{figure}

Fig.~\ref{fig:heatmap_proxy} shows cell-level monitoring quality. Before adaptation, accuracy is spatially uneven: some cells classify easily while others fail under threshold mismatch. After adaptation, the heatmap becomes more uniform across the $9 \times 9$ grid. Spatial uniformity matters because overload actions---local offload, temporary aerial coverage, or selective service restriction---should not depend on a handful of easy cells.

\FloatBarrier
\section{Discussion and Limitations}
\label{sec:discussion}

We found that baseline evaluation alone overstates emergency monitoring reliability. Diffusion-synthesized surges expose hidden threshold fragility while preserving useful ranking quality; adaptation largely restores usable alerts at $\tau=0.5$, with remaining seed variance.

\textbf{Practical relevance.} Operating-point drift and calibration-versus-adaptation inform \emph{network planning and scheduling} by showing when fixed alert rules must be recalibrated before overload actions. The diffusion protocol is a reusable \emph{testing platform} under disaster-like surge shift.

\textbf{Limitations and Future work.} Internet CDR is an indirect proxy with no paired RRC/NGAP trace. Labels and surges are synthetic on one Milano window; three seeds leave post-adaptation F1 std.\ $14.37\%$, so this is not a certified alert rule. Reported $\tau^*$ and $\Delta\tau$ are oracle values on evaluation grids. The \texttt{internet\_only} versus \texttt{three\_channel} comparison confounds channels with compact CNN versus ResNet50. Spatial realism is coarse (z-score/MMD$^2$, correlation). Disaster-period traces, a held-out calibration split, same-backbone ablations, cross-city transfer, and online recalibration remain future work.

\section{Conclusion}
\label{sec:conclusion}

Fixed-threshold monitoring can fail under realistic surge shift even when ranking quality stays strong; guided adaptation can restore usable cell-level alerts from coarse Internet CDR proxies at $\tau=0.5$, subject to the oracle-$\tau^*$ and seed-variance caveats above. Relative to our companion IEEE TII work on multi-scenario malicious-traffic robustness in a CPS-security setting, this paper contributes a focused emergency-monitoring view centered on threshold drift, multi-seed behavior, and lightweight single-channel evaluation.

\end{document}